\documentclass{appolb}
\usepackage{epsfig}
\usepackage{my}
\def\CASCADE{{\sc Cascade}}

\newcommand{\alphas}{\alpha_s}

\newcommand{\asb}{{\bar \alpha}_\mathrm{s}}
\newcommand{\alphasb}{\bar{\alpha}_s}

\newcommand{\kt}{k_{t}}

\newcommand{\CCFM}{CCFM}

\newcommand{\CASCADEMC}{CASCADEMC}

\begin{document}
\eqsec  
\title{Investigation of next-to-leading effects in CCFM%
\thanks{Presented at DIS 2002 Hadronic Final States - Cracow 2 May 2002}%
}
\author{H. Jung
\address{Department of Physics, Lund University,
        Sweden}
}
\maketitle
\begin{abstract}
The effect of formally next-to-leading contributions to the CCFM evolution
equation are discussed.
\end{abstract}
  
\section{Introduction}

The CCFM~\cite{\CCFM}
evolution equation in the framework of $k_t$-factorization and its practical
realization in the Monte Carlo program \CASCADE ~\cite{\CASCADEMC}
has been shown to be very successful in describing a bulk of experimental 
measurements~\cite{CASCADEMC,%
jung-hq-2001,smallx_coll_2001,Baranov2002},
which were not  described in the collinear approach. However, as BFKL, the
CCFM equation was derived in the high energy approximation keeping only the
singular terms (i.e. $1/z$ and $1/(1-z)$) in the splitting function $P_{g}$.
The question arises, whether the other terms, which are present in the DGLAP
splitting function, are already small enough to be neglected at the energies of
present colliders. Also the scale in the running $\alphas$ was originally 
treated differently for the small and large $z$ parts.
\section{Next-to-leading effects}
The splitting of $k_{i-1} \to k_i p_i$, where $k$ ($p$) are the four-momentum 
vectors of the
propagator (emitted) gluon, respectively,  
with momentum fractions $x_{i-1}$, $x_i$ and the
splitting variable $z=x_i/x_{i-1}$, is described by the splitting function
$P_g$.  
The original CCFM~\cite{\CCFM} splitting function $P_{g}$ was given by:
\begin{equation}
\tilde{P}_g(z,q^2,\kt^2)= \frac{\alphasb(q^2(1-z)^2)}{1-z} + 
\frac{\alphasb(\kt^2)}{z} \Delta_{ns}(z,q^2,\kt^2)
\label{Pgg}
\end{equation}
where $q = p_t/(1-z)$ and  the non-Sudakov form factor 
$\Delta_{ns}$ was defined as:
\begin{equation}
\log\Delta_{ns}(z,q^2,\kt^2) =  -\alphasb
                  \int_{z}^1 \frac{dz'}{z'} 
                        \int \frac{d q^2}{q^2} 
              \Theta(\kt-q)\Theta(q-z'q)
                  \label{non_sudakov}                   
\end{equation}
Here only the singular terms $1/z$ and $1/(1-z)$ were included and for
simplicity the scale in the running $\alphas$ was not treated in the same manner
for the small and large $z$ part.
\par
In the high energy approximation, the inclusion of the non-singular terms in the
splitting function $P_g$ as well as changes in the scale of $\alphas$ are
considered as next-to-leading effects. In the following we investigate the
numerical importance of these effects at present collider energies. 
\subsection{Scale of $\alpha_s$}
Due to the complicated structure of the CCFM splitting function, for simplicity
the transverse momentum of the propagator gluon,
$k_t$, was used as the scale in the running $\alphas$, whereas next-to-leading
order calculations suggest, that the proper scale is the transverse momentum of
the emitted gluon, $p_t$, for full range of $z$ (for a summary of the
arguments see \cite{smallx_coll_2001}).
\begin{figure}[htb]
  \begin{center}
  \vspace*{-1.0cm}
    \epsfig{file=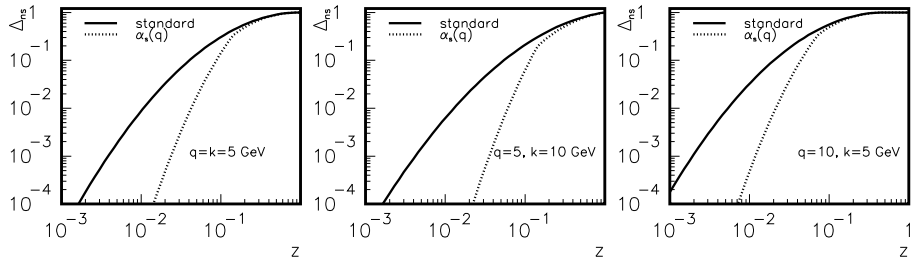,width=13cm}
  \vspace*{-0.8cm}
    \caption{{\it The non-Sudakov form factor $\Delta_{ns}$  for three
    different values of $k_t/q_t$ as a function of the splitting variable $z$
    according to eq.(\protect\ref{non_sudakov}) (solid) and 
    eq.(\ref{non_sudakov_asq}) (dotted). 
     }
    \label{deltans_alphas}}
  \vspace*{-1.0cm}
  \end{center}
\end{figure}
\par
Changing the scale in $\alphas$ from $\kt$ to $p_t$ (with 
$p_t \sim q$ for $z \to 0$), also  the non-Sudakov form factor  needs to be
properly changed, resulting in:
\begin{equation}
\tilde{P}= \frac{\alphasb(q(1-z))}{1-z} + 
           \frac{\alphasb(q)}{z} \Delta_{ns}(z,q,k_{t})
\label{Pgg_asq}
\end{equation}
\begin{equation}
\log\Delta_{ns} = -\int_0^{1} \frac{dz'}{z'}
                  \int \frac{d q'^2}{q'^2}                     
                  \alpha_s(q')\Theta(k_{t}-q')\Theta(q'-z'q)
\end{equation}
which leads to:
\begin{equation}
\log\Delta_{ns} = -\int_0^{1} \frac{dz'}{z'}
                  \int^{k_{t}^2}_{(z'q)^2} \frac{d q'^2}{q'^2}
                  \frac{1}{\log(q'/\Lambda_{QCD})}
		\label{non_sudakov_asq} 	
\end{equation}
\begin{figure}[htb]
  \begin{center}
  \vspace*{-1.0cm}
    \epsfig{file=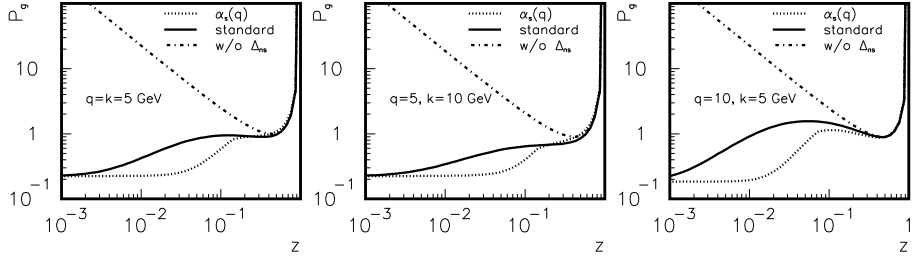,width=13cm}
  \vspace*{-0.8cm}
    \caption{{\it The splitting function $P_g$  for three
    different values of $k_t/q_t$ as a function of the splitting variable $z$
    according to eq.(\ref{Pgg},\ref{non_sudakov}) (solid)
    eq.(\ref{Pgg_asq},\ref{non_sudakov_asq}) (dotted), and with $\Delta_{ns}=1$
    (dashed)}
    \label{pgg_alphas}}
  \vspace*{-1.0cm}
  \end{center}
\end{figure}
Due to the angular ordering constraint $q'> z'q$, $q'$ can become very small
and even $q' < \Lambda_{QCD}$ at small values of $z'$. Thus a cutoff is
required.
For $ z'q < q_0 = 0.71$~GeV we fix $\alphas(q_0)=0.5$, but keep the full 
angular ordering constraint in the integral.
\par
In Fig.~\ref{deltans_alphas} we compare the new non-Sudakov form factor 
$\Delta_{ns}$ with the standard one for three different values of $k_t/q_t$. It
is interesting to note, that everywhere the very small $z$ values are
highly suppressed.
\par
In Fig.~\ref{pgg_alphas} the splitting function $P_g$ (dotted)
is plotted as a function
of $z$ for three different values of $k_t/q_t$. Also shown for comparison is the
splitting function without any suppression from the
non-Sudakov form factor ($\Delta_{ns}=1$
dashed) and the standard version of the splitting function from
eqs.(\ref{Pgg},\ref{non_sudakov}) (solid). One can clearly see how 
the different non-Sudakov form factors suppress the small $z$ region of $P_g$.
		  
\subsection{Non-singular terms in Splitting function}
Another source of next-to-leading-log corrections is the gluon splitting 
function itself.
At very high energies, the $1/z$ term in $P_{gg}$, included in BFKL and CCFM,
will certainly be dominant. However, the question is whether including 
just this term is sufficient at energies available at present colliders.
\par
The implementation of the full
DGLAP splitting function into CCFM is problematic.
Naively one would simply replace
$  \frac1{1-z} \to \frac{1}{1-z} - 2 + z(1-z)\,$ 
in the CCFM splitting function. But this can lead to negative
branching probabilities. 
\par
In \cite{smallx_coll_2001} it was suggested to use:
\vspace*{-0.2cm}
\begin{eqnarray}
P(z,q,k)& = &\asb \left(\kt^2\right) \left( \frac{(1-z)}{z} + (1-B)z(1-z)\right)
\Delta_{ns}(z,q,k) \\
& &  + \asb\left((1-z)^2q^2\right) \left(\frac{z}{1-z} + Bz(1-z)\right)
\nonumber
\label{Pgg_fullsplitt}
\end{eqnarray}
where $B$ is a parameter to be chosen arbitrarily between $0$ and $1$,
we take $B=0.5$. 
As a consequence of the replacement, the Sudakov form factor will change, but 
also the non-Sudakov form factor needs to be replaced by:
\begin{equation}
\log\Delta_{ns} =  -{\bar \alpha}_s\left(\kt^2\right)
                  \int_0^1 dz'
                        \left( \frac{1-z}{z'} + (1-B)z(1-z) \right)
                        \int \frac{d {q'}^2}{{q'}^2} 
              \Theta(k-q')\Theta(q'-z'q)
\label{non_sudakov_fullsplitt}
\end{equation}
\begin{figure}[htb]
  \begin{center}
  \vspace*{-1.0cm}
    \epsfig{file=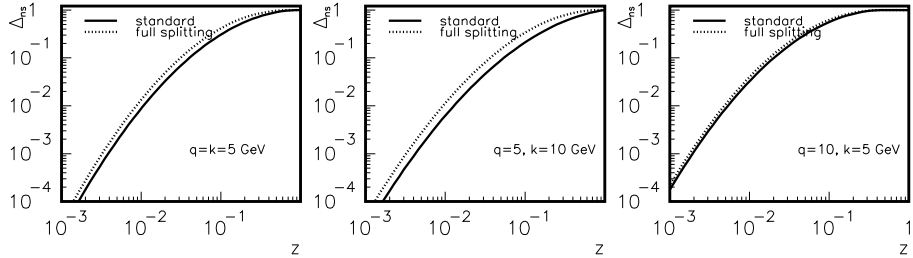,width=13cm}
  \vspace*{-0.8cm}
    \caption{{\it The non-Sudakov form factor $\Delta_{ns}$  for three
    different values of $k_t/q_t$ as a function of the splitting variable $z$
    according to eq.(\protect\ref{non_sudakov}) (solid) and 
    eq.(\ref{non_sudakov_fullsplitt}) (dotted). 
     }
    \label{deltans_fullsplitt}}
  \vspace*{-1.0cm}
  \end{center}
\end{figure}
\par
In Fig.~\ref{deltans_fullsplitt} we compare the new non-Sudakov form factor 
$\Delta_{ns}$ with the standard one for three different values of $k_t/q_t$. 
\begin{figure}[htb]
  \begin{center}
  \vspace*{-1.0cm}
    \epsfig{file=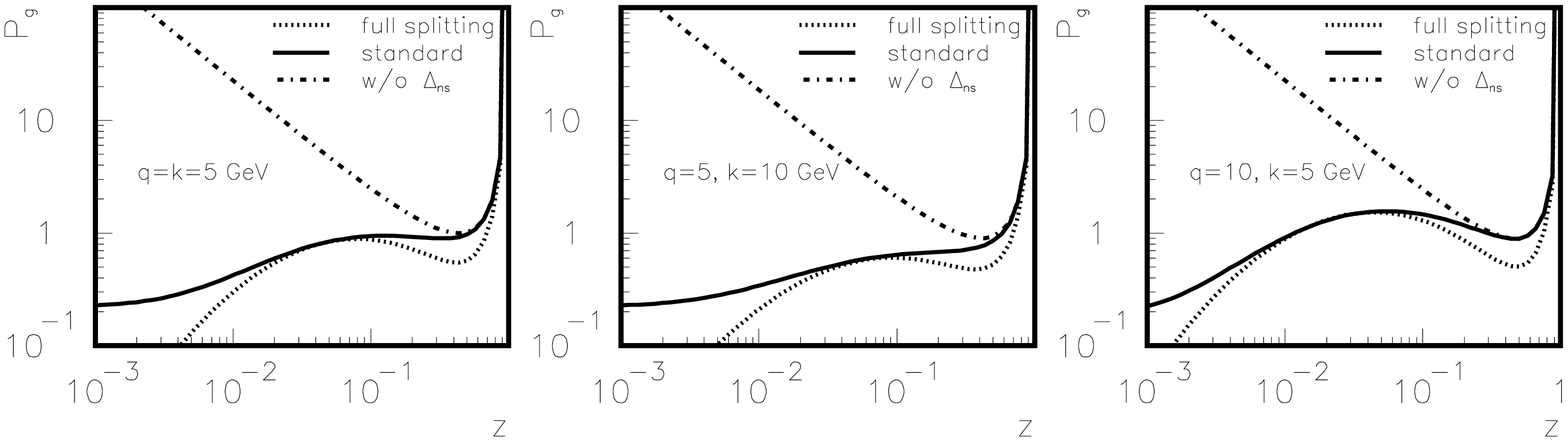,width=13cm}
  \vspace*{-0.8cm}
    \caption{{\it The splitting function $P_g$  for three
    different values of $k_t/q_t$ as a function of the splitting variable $z$
    according to eq.(\ref{Pgg},\ref{non_sudakov}) (solid)
    eq.(\ref{Pgg_fullsplitt},\ref{non_sudakov_fullsplitt}) (dotted), and with $\Delta_{ns}=1$
    (dashed)}
    \label{pgg_fullsplitt}}
  \vspace*{-0.7cm}
  \end{center}
\end{figure}
\par
In Fig.~\ref{pgg_fullsplitt} the splitting function $P_g$ (dotted)
is plotted as a  function
of $z$ for three different values of $k_t/q_t$. 
Also shown for comparison is the
splitting function without non-Sudakov form factor ($\Delta_{ns}=1$
dashed) and the standard version of the splitting function from
eqs.(\ref{Pgg},\ref{non_sudakov}) (solid). Here the effect of the different
form of the splitting function $P_g$ becomes obvious
already at values of $z \sim 0.5$, whereas the non-Sudakov
form factor is similar to the standard one. One should note that especially in
the region of medium $z$, the new branching probability (including the
non-singular terms) becomes smaller.
\subsection{Consequences for forward jet production}
\begin{figure}[h]
\begin{center}
  \vspace*{-0.5cm}
    \epsfig{file=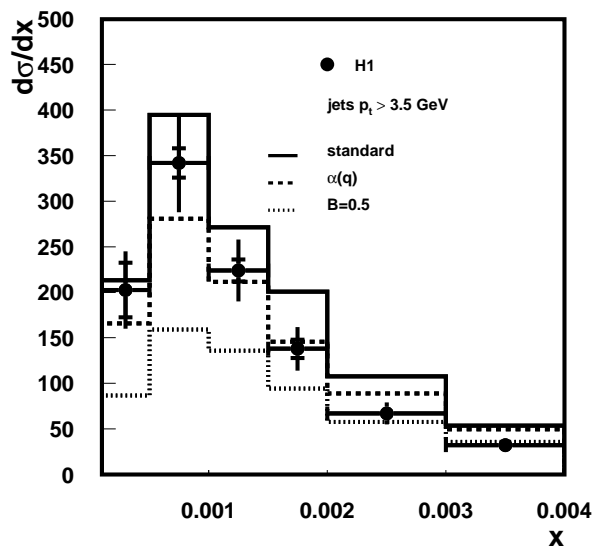,width=8.0cm}
  \vspace*{-0.2cm}
    \caption{{\it The cross section for forward jet production as a function of $x$, 
 compared to H1 data~\protect\cite{H1_fjets_data}
 }
    \label{fwdjet}}
 \end{center}
  \vspace*{-0.4cm}
\end{figure}

In Fig.~\ref{fwdjet} we show the predictions for forward jet production at 
HERA~\cite{H1_fjets_data} for the different scenarios discussed above.
 All cases have been re-fitted to the structure function $F_2$, with a
similarly good $\chi^2/ndf$. It becomes obvious, that the prediction for forward
jet production is rather sensitive to the details of the gluon splitting
function.
\vspace*{-1.0cm}
\section*{Acknowledgments}
This paper is dedicated to the memory of Bo Andersson, who died 
unexpectedly from a heart attack on March 4th, 2002. 
I have learned so much from him.
I am very grateful to G.~Salam for all his ideas and advice    
concerning CCFM and the next-to-leading contributions, which formed the 
basis for this contribution. 
\raggedright

\end{document}